%% file: Article.tex
\documentclass[review,onefignum,onetabnum]{siamart171218}
\usepackage[a4paper,margin=1in]{geometry}
\usepackage{listings}
\usepackage{xcolor} 
\usepackage{placeins}
\usepackage{graphicx}
\usepackage{subcaption}

\input{ex_shared}

\ifpdf
\hypersetup{
  pdftitle={An Example Article},
  pdfauthor={D. Doe, P. T. Frank, and J. E. Smith}
}
\fi

\myexternaldocument{ex_supplement}

\begin{document}

\maketitle

\begin{abstract}
In recent years, a surge in the popularity of trap music among adolescents has prompted concerns about its potential influence on youth behavior and educational outcomes. In this study, we develop a novel compartmental model using a system of differential equations to explore the relationship between exposure to trap music and school dropout rates among Costa Rican adolescents aged 13 to 17. The model divides the population into distinct compartments representing susceptible individuals, casual listeners, active participants, those exhibiting risk-associated behaviors, and ultimately, school dropouts. Key parameters, including transmission via peer influence, progression rates between exposure stages, and recovery dynamics, capture the complex interplay between cultural diffusion and behavioral change. Analytical investigation of the {\it basic reproductive number}, $R_0$, and both trap-free and endemic equilibrium states provide insight into the conditions under which the influence of trap music proliferates. Numerical simulations, implemented in MATLAB, further illustrate how parameter variations, especially the potential for recovery, affect the system’s dynamics. Our results suggest that although exposure to trap music is widespread, the progression to adverse behavioral outcomes leading to school dropout is highly sensitive to intervention strategies, offering valuable implications for educational policy and targeted preventive measures.
\end{abstract}




\section{Introduction}

In recent decades, the evolution of popular music has increasingly intersected with social and cultural dynamics, with trap music emerging as a particularly influential genre among adolescents. Originating in the early 1990s in the southern United States, trap music initially reflected the harsh realities of urban life. Its distinctive beats, lyrical focus on street culture, and themes of hardship and survival characterized it. Over time, the genre has evolved, blending elements of hip-hop, rap, crunk, and electronic music, and has subsequently transcended geographic and cultural boundaries to become a global phenomenon~\cite{Ifema,Estilo}. In Latin America, for instance, the genre has been popularized by leading artists such as Bad Bunny, Anuel AA, Cazzu, and Neo Pistea, whose works have not only redefined the musical landscape but have also sparked debates regarding the sociocultural implications of trap music.

The thematic content of trap music, which often includes the glorification of violence, drug use, and a hyper-masculine portrayal of gender roles, has generated significant academic and public concern. Critics argue that these representations may serve as behavioral models for impressionable youth, potentially normalizing risky or deviant behavior. In many music videos and lyrical narratives, trap music frequently incorporates symbolic imagery, ranging from religious iconography to representations of wealth and excess, which may further reinforce consumerist and nonconformist attitudes~\cite{Laura,Ruiz,Esther}. Such aesthetic and thematic elements have raised questions about trap music's potential role in shaping its listeners' identity and behavior, particularly during the critical developmental stage of adolescence.

Adolescence is widely recognized as a formative period marked by exploring identity, developing social relationships, and a heightened sensitivity to external influences. Psychological theories, including those stemming from psychoanalytic and developmental frameworks, have long underscored the vulnerability of adolescents to cultural and peer influences~\cite{Carretero}. As teenagers navigate the complexities of identity formation and social integration, they are continuously exposed to a barrage of information via traditional media and digital platforms such as social networks, streaming services, and video-sharing websites. Within this intricate network of influences, music emerges as a potent medium for socialization, offering both a source of emotional expression and a template for behavior that can be readily imitated.

Costa Rica presents a fascinating context for examining these dynamics. Despite the nation’s notable educational achievements, recent reports indicate that school dropout rates among adolescents have reached alarming levels, with over 21,000 students leaving the education system as of September 2023~\cite{Semanario,Maria}. While various factors, including economic hardship, family dynamics, and broader social challenges, are known to contribute to school dropout, the potential impact of cultural phenomena such as trap music has received comparatively less scholarly attention. Given that a significant portion of the trap music audience comprises young individuals, it is plausible that the genre’s distinctive messaging and associated lifestyle portrayals may interact with existing social vulnerabilities, influencing academic engagement and persistence.

Against this backdrop, the present study seeks to explore the influence of trap music on school dropout rates among Costa Rican adolescents aged 13 to 17. Recognizing that behavioral change is a complex process driven by a confluence of factors, we propose a novel compartmental model that captures the progression from mere exposure to trap music to adopting risky behaviors and, ultimately, school dropout. The model divides the adolescent population into discrete compartments—those susceptible to trap music, casual listeners who remain unaffected, individuals exposed to and actively engaged with the genre, and those who exhibit risk-associated behaviors culminating in school dropout. In doing so, the model provides a structured framework to quantify the transmission dynamics of cultural influence in a manner analogous to epidemiological models successfully applied to the study of rumor propagation and behavior diffusion~\cite{Centola,Isea,Lopez}.

The model is parameterized to reflect the various rates at which adolescents transition between these states, incorporating factors such as peer influence, the intrinsic appeal of trap music, the effectiveness of countervailing social or educational interventions, and the potential for behavioral ''recovery" from risky engagement. By deriving and analyzing the basic reproductive number, $R_0$, and the conditions for trap-free and endemic equilibria, our analytical framework provides insight into the thresholds and dynamics that govern the proliferation of adverse behaviors linked to trap music exposure. Furthermore, numerical simulations implemented in MATLAB allow for the exploration of various scenarios, offering a means to assess the system's sensitivity to changes in key parameters, such as the rate of re-exposure among adolescents who have previously disengaged from the genre.

The current study contributes to a deeper understanding of how contemporary cultural phenomena can influence educational outcomes through this interdisciplinary approach, which blends sociocultural theory with rigorous mathematical modeling. The insights derived from this research have the potential to inform both policymakers and educators by highlighting the need for targeted interventions that address not only the economic and social determinants of school dropout but also the cultural and media-related factors that may exacerbate these challenges.

The remainder of the paper is organized as follows. Section~\ref{sec:model} presents a detailed mathematical model description, including its underlying assumptions and the formulation of the system's differential equations. Section~\ref{sec:experiments} discusses the numerical simulations. Finally, Section~\ref{sec:conclusions} offers concluding remarks and directions for future research.

\section{Mathematical Model}\label{sec:model}

In this study, the total population \(N\) represents Costa Rican adolescents aged 13 to 17 and is assumed to remain constant over time. This population is divided into several compartments based on their interaction with trap music and the resulting behavioral outcomes. Specifically, \(S(t)\) denotes the number of susceptible individuals who have not yet been exposed to trap music. Among those exposed to cultural influences, \(C(t)\) represents adolescents who, despite being in an environment where trap music is prevalent, opt not to engage with it. The compartment \(E(t)\) consists of individuals exposed to trap music and at risk of transitioning to active listening behaviors.

Within the active listening group, the model further differentiates between two subgroups. \(T_0(t)\) corresponds to regular listeners of trap music who do not exhibit any risky or adverse behaviors, whereas \(T_1(t)\) identifies those listeners who, in addition to their engagement with the music, adopt behaviors that significantly increase their risk of school dropout. The state \(D(t)\) captures the number of adolescents who have dropped out of school due to such behaviors. Finally, \(R(t)\) comprises individuals who have disengaged from trap music, including those who have recovered from risky engagement and, in some cases, dropouts who no longer listen to the music.

The following is the system of differential equations:
\begin{align*}
\frac{dS}{dt} &= \mu N - \beta S \frac{p\,T_0 + q\,T_1 + r\,D}{N} - (\mu+\alpha) S,\\
\frac{dC}{dt} &= \alpha S + \phi E - \mu C, \\
\frac{dE}{dt} &= \beta S \frac{p\,T_0 + q\,T_1 + r\,D}{N} - (\mu+\phi+\xi) E, \\
\frac{dT_0}{dt} &= \xi E - (\mu+\gamma_0+\omega) T_0, \\
\frac{dT_1}{dt} &= \omega T_0 + \rho R \frac{p\,T_0 + q\,T_1 + r\,D}{N} - (\mu+\gamma_1+\delta) T_1, \\
\frac{dR}{dt} &= \gamma_0 T_0 + \gamma_1 T_1 + \gamma_d D - \rho R \frac{p\,T_0 + q\,T_1 + r\,D}{N} - \mu R, \\
\frac{dD}{dt} &= \delta T_1 - (\mu+\gamma_d) D. \label{eq:D}
\end{align*}

Here, the parameter \(\mu\) denotes the natural exit rate (or population turnover), while \(\beta\) is the transmission rate capturing peer influence. The parameter \(\alpha\) represents the rate at which susceptible individuals choose not to engage with trap music, leading them into the \(C\) compartment. The rate \(\phi\) measures the shift from exposure to non-engagement, and \(\xi\) is the rate at which exposure results in active listening without immediate risk, transitioning individuals from \(E\) to \(T_0\). The parameter \(\omega\) governs the progression from safe active listening (\(T_0\)) to risky behaviors (\(T_1\)), and \(\delta\) is the rate at which risky behavior leads to school dropout (\(D\)). Additionally, \(\gamma_0\), \(\gamma_1\), and \(\gamma_d\) denote the rates at which individuals in compartments \(T_0\), \(T_1\), and \(D\), respectively, move to the recovered state \(R\). Finally, \(\rho\) is the rate at which recovered individuals revert to active listening, and the weights \(p\), \(q\), and \(r\) quantify the relative influence of the \(T_0\), \(T_1\), and \(D\) compartments on the transmission dynamics.

This mathematical model provides a structured framework to explore the progression from initial exposure to trap music to the eventual risk of school dropout, thereby enabling the analysis of intervention strategies to mitigate adolescent adverse outcomes.

\section{Results}
\subsection{Basic Reproductive Number}

The basic reproductive number \(R_0\) is given by (details can be viewed in \ref{R0}):

\[
R_0 = \frac{S \beta p}{N (\mu + \phi + \xi)} + \frac{S \beta q \xi}{N (\mu + \gamma_0 + \omega)(\mu + \phi + \xi)} + \frac{S \beta \omega r \xi}{N (\mu + \gamma_1 + \delta)(\mu + \gamma_0 + \omega)(\mu + \phi + \xi)}
\]

\subsection{Trap-Free Equilibrium Point} A possible equilibrium state for this \linebreak model is the diffusion-free state, given by

$$(S^*,C^*,E,T_0,T_1,R,D)=\left(\displaystyle \frac{\mu N}{\mu+\alpha},\frac{\alpha N}{\mu+\alpha},0,0,0,0,0\right)$$ 

This equilibrium point is determined solely by the entry and exit rates $\mu$ of individuals in the population being studied, as well as the classes $S$ and $C$  and the transmission parameter between them.

\subsection{Endemic Equilibrium}
Using the system of differential equations, we have:
$$S^*=\displaystyle\frac{\mu N}{\frac{\beta}{N}(pT_0^*+qT_1^*+rD^*)+\mu+\alpha}$$            

it follows that,
$$C^*=\displaystyle \frac{\alpha}{\mu}S^*+\frac{\phi}{\mu}E^*.$$

Next, we have:
$$E^*=\beta S^*\frac{pT^*_0+qT^*_1+rD^*}{N}$$

$$T^*_0=\frac{\xi E^*}{\mu+\gamma_0+\omega}$$
and
$$T^*_1=\frac{\omega}{\mu+\gamma_1+\delta}T^*_0+\rho$$
$$R^*\frac{pT^*_0+qT^*_1+rD^*}{(\mu+\gamma_1+\delta)N}$$

\noindent Finally,
$$R^*=\frac{\gamma_0 T^*_0 + \gamma_1 T^*_1 + \gamma_d D^*}{\rho  \frac{pT^*_0+qT^*_1+rD^*}{N} + \mu}$$
and
$$D^*=\frac{\delta T^*_1}{\mu+\gamma_d}.$$

Simplifying, we have
$$pT^*_0+qT^*_1+rD^*=\frac{\xi E^*}{\mu+\gamma_0+\omega}\left(p+\frac{q\omega}{\mu+\gamma_1+\delta}+\frac{r\omega\delta}{(\mu+\gamma_d)(\mu+\gamma_1+\delta)}\right)$$

Now, 
$$ \frac{\beta S^*}{N}=\frac{\mu \beta N}{\beta(pT_0^*+qT_1^*+rD^*)+\mu N+\alpha N}$$

$$\frac{\beta S^*}{N}=\frac{\mu \beta N}{\beta\cdot\frac{\xi E^*}{\mu+\gamma_0+\omega}\left(p+\frac{q\omega}{\mu+\gamma_1+\delta}+\frac{r\omega\delta}{(\mu+\gamma_d)(\mu+\gamma_1+\delta)}\right)+\mu N+\alpha N}.$$  

Furthermore,
$$A=\frac{\xi }{\mu+\gamma_0+\omega}\left(p+\frac{q\omega}{\mu+\gamma_1+\delta}+\frac{r\omega\delta}{(\mu+\gamma_d)(\mu+\gamma_1+\delta)}\right)$$
and
$$B=\frac{\beta\xi }{\mu+\gamma_0+\omega}\left(p+\frac{q\omega}{\mu+\gamma_1+\delta}+\frac{r\omega\delta}{(\mu+\gamma_d)(\mu+\gamma_1+\delta)}\right).$$

Then:
$$E^*=N\left(\mu-\left(\frac{\mu + \alpha}{\frac{\beta \xi }{\mu+\gamma_0+\omega}\left(p+\frac{q\omega}{\mu+\gamma_1+\delta}+\frac{r\omega\delta}{(\mu+\gamma_d)(\mu+\gamma_1+\delta)}\right)}\right)\right)$$

$$S^*=\frac{N}{\frac{\beta\xi}{\mu+\gamma_0+\omega}\left(p+\frac{q\omega}{\mu+\gamma_1+\delta}+\frac{r\omega\delta}{(\mu+\gamma_d)(\mu+\gamma_1+\delta)}\right)}$$

$$C^*=\frac{N\left(\alpha(1-\phi)+\phi \left(\frac{\beta\xi }{\mu+\gamma_0+\omega}\left(p+\frac{q\omega}{\mu+\gamma_1+\delta}+\frac{r\omega\delta}{(\mu+\gamma_d)(\mu+\gamma_1+\delta)}\right)- \mu\right)\right)}{\mu \cdot \frac{\beta\xi }{\mu+\gamma_0+\omega}\left(p+\frac{q\omega}{\mu+\gamma_1+\delta}+\frac{r\omega\delta}{(\mu+\gamma_d)(\mu+\gamma_1+\delta)}\right)}$$
 
$$T^*_0=\frac{N\xi\mu}{\mu+\gamma_0+\omega}-\frac{N(\mu+\alpha)}{\beta \left(p+\frac{q\omega}{\mu+\gamma_{1}+\delta}+\frac{r\omega \delta}{(\mu+\gamma_d)(\mu+\gamma_{1}+\delta)}\right)}$$

$$T^*_1=\frac{N\omega}{\mu+\gamma_1+\delta}\left(\frac{\xi\mu}{\mu+\gamma_0+\omega}-\frac{(\mu+\alpha)}{\beta \left(p+\frac{q\omega}{\mu+\gamma_{1}+\delta}+\frac{r\omega \delta}{(\mu+\gamma_d)(\mu+\gamma_{1}+\delta)}\right)}\right)$$

for $D^*$, we have:
$$ D^*= \frac{ \omega N}{(\mu+\gamma_d)(\mu+\gamma_1+\delta)}\left(\frac{\xi\mu}{\mu+\gamma_0+\omega}-\frac{(\mu+\alpha)}{\beta \left(p+\frac{q\omega}{\mu+\gamma_{1}+\delta}+\frac{r\omega \delta}{(\mu+\gamma_d)(\mu+\gamma_{1}+\delta)}\right)}\right).$$

And finally:

\[
R^{*}_{0}=\frac{N}{\mu} \left(\frac{\xi \mu}{\mu+\gamma_0+\omega} 
- \frac{\mu+\alpha}{\beta \left(p+\frac{q\omega}{\mu+\gamma_1+\delta} + \frac{r\omega \delta}{(\mu+\gamma_d)(\mu+\gamma_1+\delta)}\right)}\right) 
\left(\gamma_0 + \frac{\gamma_1 \omega}{\mu+\gamma_1+\delta} 
+ \frac{\gamma_d \delta}{(\mu+\gamma_d)(\mu+\gamma_1+\delta)}\right).
\]

\section{Numerical Simulations}\label{sec:experiments}







In our numerical simulations, we tracked the evolution of three key population compartments, $T_0$ (the blue curve for regular trap music listeners), $T_1$ (the orange curve for those adopting riskier behaviors), and D (the yellow curve representing school dropouts)—over time in a population of 10,000 adolescents. In the first scenario, with the recovery parameter set to $\rho=0$, individuals who moved to the recovered state never re-enter the listening cycle. Under these conditions, the $T_0$ compartment exhibits a rapid initial rise before reaching a steady value (around 60, 40, 45, or 20 depending on the specific values of the influence weights p, q, and r), suggesting that trap music quickly becomes a habitual background soundtrack for many. In contrast, the $T_1$ compartment also grows fast initially. Still, it stabilizes at significantly lower levels (roughly 8, 5, 5, and 3), indicating that only a tiny fraction of adolescents transition into risky behavior—likely due to social resistance, educational influences, or family support.

Meanwhile, the dropout compartment $D$ follows a similar trend to $T_0$ but saturates at even lower numbers (approximately 50, 30, 35, and 20), implying that while some adolescents eventually drop out, the overall impact is relatively contained. In the second scenario, when $\rho$ is nonzero—meaning individuals in the recovered state can rejoin $T_0$ and restart the cycle—the dynamics shift dramatically. Although the $T_0$ and $T_1$ compartments retain their early growth patterns, the possibility of re-engagement leads to a pronounced increase in the dropout population, with D reaching very high values (around 700, 1700, 1300, and 1100), effectively illustrating a vicious cycle where repeated exposure escalates the risk of school dropout. In simple terms, the simulations highlight that while casual exposure to trap music might not spell immediate disaster, allowing for repeated engagement can set off a chain reaction with severe educational consequences—a wake-up call for any policymaker or educator.

\begin{figure}[h!]
    \centering
    \begin{subfigure}[b]{0.45\textwidth}
        \includegraphics[width=\textwidth]{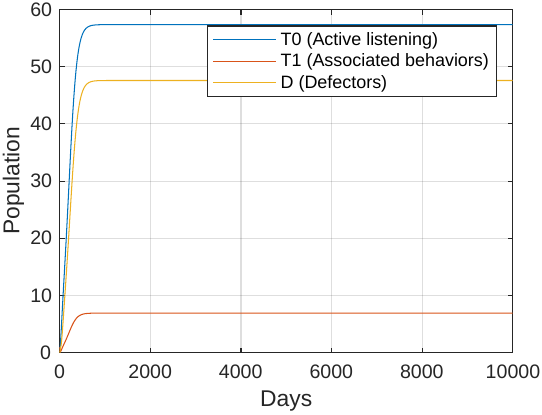}
        \caption{$p=0.8$, $q=0.5$, $r=0.7$, $R_{0}=4.9156$}
    \end{subfigure}
    \hfill
    \begin{subfigure}[b]{0.45\textwidth}
        \includegraphics[width=\textwidth]{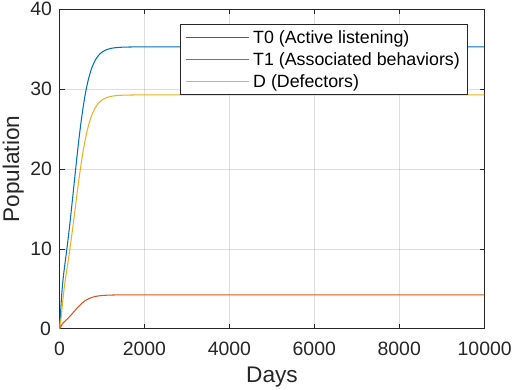}
        \caption{$p=0.7$, $q=0.3$, $r=0.7$, $R_{0}=4.2933$}
    \end{subfigure}

    \vskip\baselineskip
    
    \begin{subfigure}[b]{0.45\textwidth}
        \includegraphics[width=\textwidth]{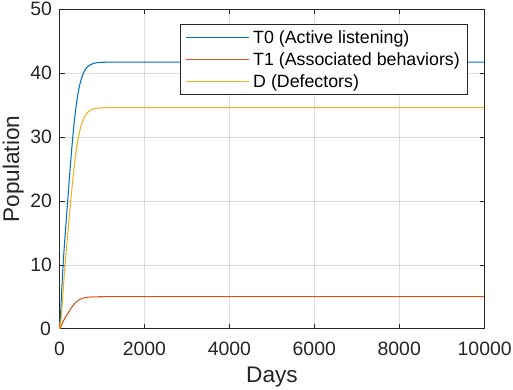}
        \caption{$p=0.9$, $q=0.3$, $r=0.5$, $R_{0}=5.5151$}
    \end{subfigure}
    \hfill
    \begin{subfigure}[b]{0.45\textwidth}
        \includegraphics[width=\textwidth]{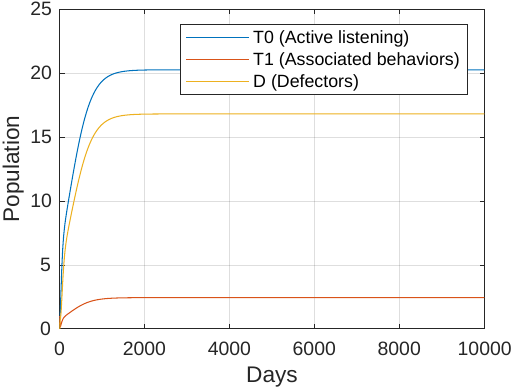}
        \caption{$p=0.9$, $q=0.1$, $r=0.4$, $R_{0}=5.5037$}
    \end{subfigure}
    \caption{Comparison of simulations for different values of parameters $p$, $q$, and $r$ when $\rho=0$.}
    \label{fig:simulations}
\end{figure}


The population at $T_0$ increases rapidly at the beginning; subsequently, it reaches a constant value (about $60, 40, 45,$ and $20$ respectively) so that the number of teenagers in $T_0$ reaches equilibrium as time passes. The behavior observed in the graphs for these curves could be associated with the diffusion of music in the susceptible population, where it could be noted that the population adopts music as a routine.

On the other hand, the curve representing $T_1$ shows rapid growth at the beginning; however, unlike $T_0$, it reaches the constant at much lower levels (close to 8, 5, 5, and 3, respectively). The above suggests that only a tiny part of the population adopts behaviors associated with trap music. This could be due to social resistance, education, or family environment.

The population $D$ (the deserter population) shows a similar pattern to $T_0$ but saturates at lower levels (near 50, 30, 35, and 20). At that time, a small percentage of the population (considering the sample of 10,000) ends up in a state of school dropout. 

\subsection{Case when $\rho\neq 0$}

\begin{figure}[h!]
    \centering
    \begin{subfigure}[b]{0.45\textwidth}
        \includegraphics[width=\textwidth]{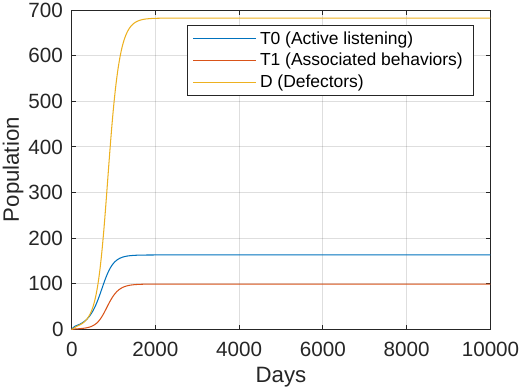}
        \caption{$\rho=1$, $p=0.9$, $q=0.1$, $r=0.4$, $R_{0}=5.5037$}
    \end{subfigure}
    \hfill
    \begin{subfigure}[b]{0.45\textwidth}
        \includegraphics[width=\textwidth]{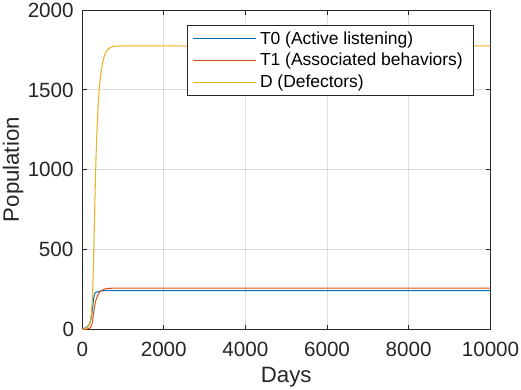}
        \caption{$\rho=2$, $p=0.7$, $q=0.5$, $r=0.8$, $R_{0}=4.3047$}
    \end{subfigure}

    \vskip\baselineskip
    
    \begin{subfigure}[b]{0.45\textwidth}
        \includegraphics[width=\textwidth]{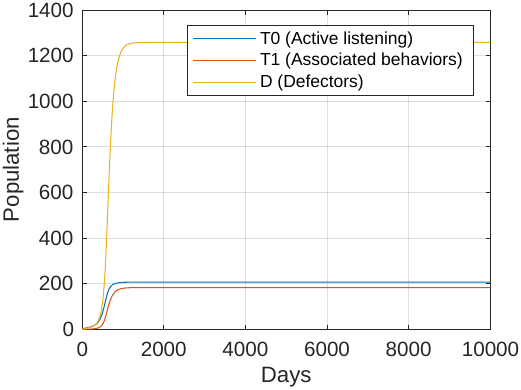}
        \caption{$\rho=1.5$, $p=0.8$, $q=0.2$, $r=0.5$, $R_{0}=4.8985$}
    \end{subfigure}
    \hfill
    \begin{subfigure}[b]{0.45\textwidth}
        \includegraphics[width=\textwidth]{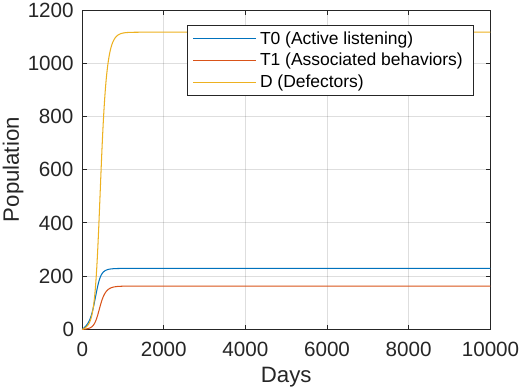}
        \caption{$\rho=0.5$, $p=0.7$, $q=0.5$, $r=0.9$, $R_{0}=4.3048$}
    \end{subfigure}

    \caption{Comparison of simulations for different values of parameters $p$, $q$, and $r$ when $\rho \neq 0$.}
    \label{fig:simulations_rho}
\end{figure}






\section{Impact of Recovery on Population Dynamics}

In the numerical simulations presented, various values of $\rho$ were explored to assess the impact of recovery dynamics on the system's long-term behavior. The parameter $\rho$ represents the rate at which individuals in the recovered compartment ($R$) re-engage with trap music, effectively transitioning back into the susceptible $T_0$ category. The results indicate that when $\rho$ is nonzero, a feedback loop emerges where individuals who initially disengaged from trap music are reintegrated into the listening population, thereby perpetuating the cycle of influence.

From a behavioral perspective, this suggests that individuals who previously ceased listening to trap music—whether due to external interventions, personal choices, or other deterrents—can be drawn back into the cycle, potentially due to factors such as social pressure, peer influence, nostalgia, or continued exposure through digital media platforms. This cyclical nature of engagement aligns with existing sociological theories on cultural diffusion and recidivism in behavioral adoption, where temporary disengagement does not necessarily equate to permanent detachment from a particular cultural or social phenomenon.

The significance of this dynamic is particularly evident in the dropout population ($D$). In simulations where $\rho = 0$, meaning that recovered individuals do not return to the system, the dropout population stabilizes at comparatively lower levels. However, as soon as $\rho$ takes on nonzero values, a dramatic escalation in the deserter population is observed. The number of dropouts ($D$) increases at a significantly faster rate, surpassing the steady-state levels of both $T_0$ and $T_1$. This is an important observation, as it implies that the possibility of re-engagement with trap music not only affects the size of the susceptible population but also amplifies the long-term risk of school dropout.

Quantitatively, the simulations reveal that under different values of $\rho$, the dropout population experiences a sharp and sustained increase, with final values reaching approximately 700, 1700, 1300, and 1100 individuals, depending on the specific parameter configurations. This trend suggests that even a moderate rate of re-engagement can have a compounding effect, leading to a disproportionately large impact on the dropout rate over time. Such findings underscore the necessity of intervention strategies that not only prevent initial exposure but also address potential relapse into the behavior, particularly through sustained educational and social support mechanisms.

In conclusion, the role of $\rho$ in the model highlights the importance of understanding behavioral recidivism within cultural influences. The findings suggest that one-time interventions may be insufficient in mitigating the long-term effects of trap music exposure, particularly if there are no mechanisms in place to prevent re-engagement. Future studies could further explore how different intervention strategies—such as social campaigns, media literacy education, or targeted behavioral programs—might alter the trajectory of the system, potentially reducing the magnitude of the dropout population even when $\rho$ is nonzero.

\section{Conclusions}\label{sec:conclusions}



This study presents a novel mathematical framework to analyze the potential influence of trap music on adolescent behavior and school dropout rates in Costa Rica. Using a compartmental differential equation model, we examined the transmission dynamics of cultural exposure and behavioral progression, drawing parallels to epidemiological models of social influence. Our findings offer valuable insights into the mechanisms by which cultural phenomena, such as music, can impact educational outcomes.

From the mathematical analysis, we derived the {\it basic reproductive number}, $R_0$, which serves as a key threshold for determining whether the influence of trap music propagates or diminishes within the adolescent population. The existence of both a trap-free equilibrium and an endemic equilibrium highlights the conditions under which exposure to trap music may lead to widespread adoption and associated risk behaviors. Furthermore, numerical simulations revealed that while many adolescents engage with trap music at a casual level, only a subset progresses to behaviors that significantly increase the likelihood of school dropout.

The simulations under different parameter scenarios demonstrate that peer influence plays a crucial role in the diffusion of trap music consumption and its associated behavioral consequences. Notably, when the recovery parameter ($\rho$) is set to zero, indicating that individuals do not return to the listening population, the dropout rate stabilizes at relatively lower levels. However, when $\rho$ is nonzero, allowing for repeated engagement with the genre, the model predicts a substantial increase in school dropout rates, reinforcing the importance of intervention strategies to disengage at-risk individuals from repeated exposure.

The model’s findings have crucial implications for educational policy and public health. First, the system's sensitivity to peer influence suggests that school-based interventions emphasizing positive peer interactions and alternative cultural engagement could mitigate the adverse effects of trap music exposure. Second, the model underscores the necessity of addressing the social determinants of education, including family environment and media consumption habits, to create a more comprehensive approach to dropout prevention. Lastly, our simulations' potential for behavioral recovery highlights the value of targeted rehabilitation programs that encourage disengagement from high-risk influences and re-engagement with formal education.

While this study provides a foundational framework for understanding the intersection of cultural phenomena and educational trajectories, several areas warrant further investigation. Future research should incorporate empirical data on adolescent listening habits and behavioral outcomes to refine parameter estimates and validate model predictions. Additionally, extending the model to account for socioeconomic factors and digital media consumption patterns could provide a more nuanced understanding of the drivers of school dropout.

In conclusion, our study demonstrates that mathematical modeling can be a powerful tool for exploring the social and cultural dynamics that shape adolescent behavior. By integrating analytical and computational approaches, we offer a structured methodology for assessing the impact of contemporary artistic trends on educational outcomes, ultimately informing policy decisions to foster youth well-being and academic success.

\appendix

\section{Basic Reproductive Number $R_0$}\label{R0}

Calculation of the basic reproductive number \( R_0 \) with the next generation operator:

\textbf{Step 1: Identification of infected compartments}

Given the following system of differential equations:

\[
\frac{dS}{dt} = \mu N - \frac{\beta S (p T_0 + q T_1 + r D)}{N} - (\mu + \alpha) S
\]
\[
\frac{dC}{dt} = \alpha S + \phi E - \mu C
\]
\[
\frac{dE}{dt} = \frac{\beta S (p T_0 + q T_1 + r D)}{N} - (\mu + \phi + \xi) E
\]
\[
\frac{dT_0}{dt} = \xi E - (\mu + \gamma_0 + \omega) T_0
\]
\[
\frac{dT_1}{dt} = \omega T_0 + \frac{\rho R (p T_0 + q T_1 + r D)}{N} - (\mu + \gamma_1 + \delta) T_1
\]
\[
\frac{dR}{dt} = \gamma_0 T_0 + \gamma_1 T_1 + \gamma_d D - \frac{\rho R (p T_0 + q T_1 + r D)}{N} - \mu R
\]
\[
\frac{dD}{dt} = \gamma_1 T_1 - (\mu + \gamma_d) D
\]

We identify the infected compartments as \( E \), \( T_0 \), y \( T_1 \).

\textbf{Step 2: Transmission Matrix \( \mathcal{F} \)}

The transmission matrix is:

\[
\mathcal{F} =
\begin{pmatrix}
\frac{\beta S p}{N} & \frac{\beta S q}{N} & \frac{\beta S r}{N} \\
0 & 0 & 0 \\
0 & 0 & 0
\end{pmatrix}
\]

\textbf{Step 3: Transition Matrix \( \mathcal{V} \)}

The transition matrix is:
\[
\mathcal{V} =
\begin{pmatrix}
\mu + \phi + \xi & 0 & 0 \\
-\xi & \mu + \gamma_0 + \omega & 0 \\
0 & -\omega & \mu + \gamma_1 + \delta
\end{pmatrix}
\]

\textbf{Calculation of the eigenvalues of \( \mathcal{F} \mathcal{V}^{-1} \)}

The eigenvalues of the matrix \( \mathcal{F} \mathcal{V}^{-1} \) are:
\[
\lambda_1 = \frac{S \beta p}{N (\mu + \phi + \xi)} + \frac{S \beta q \xi}{N (\mu + \gamma_0 + \omega)(\mu + \phi + \xi)} + \frac{S \beta \omega r \xi}{N (\mu + \gamma_1 + \delta)(\mu + \gamma_0 + \omega)(\mu + \phi + \xi)}
\]

\[
\lambda_2 = 0
\]

\[
\lambda_3 = 0
\]

The basic reproductive number \( R_0 \) is the largest eigenvalue, which is \( \lambda_1 \). Therefore, the basic reproductive number \(R_0\) is given by:

\[
R_0 = \frac{S \beta p}{N (\mu + \phi + \xi)} + \frac{S \beta q \xi}{N (\mu + \gamma_0 + \omega)(\mu + \phi + \xi)} + \frac{S \beta \omega r \xi}{N (\mu + \gamma_1 + \delta)(\mu + \gamma_0 + \omega)(\mu + \phi + \xi)}
\]

\section{Endemic Equilibrium Formulas} 
\label{Endemic Equilibrium Formulas}
\begin{eqnarray*}
    \frac{dS}{dt}   &=& \mu N - \beta S \frac{pT_0+qT_1+rD}{N} - (\mu+\alpha)S \\
         \beta S \frac{pT_0+qT_1+rD}{N} - (\mu+\alpha)S &=& \mu N  \\
    S&=&\displaystyle\frac{\mu N}{\frac{\beta}{N}(pT_0+qT_1+rD)+\mu+\alpha}    
         \end{eqnarray*}
         Then,
         \begin{eqnarray}\label{S*}
    S^*&=&\displaystyle\frac{\mu N}{\frac{\beta}{N}(pT_0^*+qT_1^*+rD^*)+\mu+\alpha}            
         \end{eqnarray}
   
\begin{eqnarray*}
    \displaystyle \frac{dC}{dt}&=& \alpha S + \phi E - \mu C\\
    0&=& \alpha S + \phi E - \mu C\\
    \mu C&=& \alpha S + \phi E \\
    C&=&\displaystyle \frac{\alpha}{\mu}S+\frac{\phi}{\mu}E
\end{eqnarray*}
Then,
\begin{eqnarray}\label{C*}
    C^*&=&\displaystyle \frac{\alpha}{\mu}S^*+\frac{\phi}{\mu}E^*
\end{eqnarray}

\begin{eqnarray*}
    \frac{dE}{dt}   &=& \beta S \frac{pT_0+qT_1+rD}{N} - (\mu+\phi+\xi)E\\
    (\mu+\phi+\xi)E&=&\beta S\frac{pT_0+qT_1+rD}{N}\\
    E&=&\beta S\frac{pT_0+qT_1+rD}{N}
\end{eqnarray*}
Then,
\begin{eqnarray}\label{E*}
     E^*&=&\beta S^*\frac{pT^*_0+qT^*_1+rD^*}{N}
\end{eqnarray}
\begin{eqnarray*}
    \frac{dT_0}{dt} &=& \xi E - (\mu+\gamma_0+\omega)T_0\\
    T_0&=&\frac{\xi E}{\mu+\gamma_0+\omega}
\end{eqnarray*}
Then,
\begin{eqnarray}\label{T0*}
    T^*_0&=&\frac{\xi E^*}{\mu+\gamma_0+\omega}
\end{eqnarray}

\begin{eqnarray*}
    \frac{dT_1}{dt} &=& \omega T_0 + \rho R \frac{pT_0+qT_1+rD}{N} - (\mu+\gamma_1+\delta)T_1\\
(\mu+\gamma_1+\delta)T_1&=&\omega T_0 + \rho R \frac{pT_0+qT_1+rD}{N} \\
    T_1&=&\frac{\omega}{\mu+\gamma_1+\delta}T_0+\rho R\frac{pT_0+qT_1+rD}{(\mu+\gamma_1+\delta)N}
\end{eqnarray*}
Then,
\begin{eqnarray}\label{T1*}
    T^*_1&=&\frac{\omega}{\mu+\gamma_1+\delta}T^*_0+\rho R^*\frac{pT^*_0+qT^*_1+rD^*}{(\mu+\gamma_1+\delta)N}
\end{eqnarray}

\begin{eqnarray*}
    \frac{dR}{dt}   &=& \gamma_0 T_0 + \gamma_1 T_1 + \gamma_d D - \rho R \frac{pT_0+qT_1+rD}{N} - \mu R\\
    \rho R \frac{pT_0+qT_1+rD}{N} + \mu R&=&\gamma_0 T_0 + \gamma_1 T_1 + \gamma_d D\\
     R\left(\rho  \frac{pT_0+qT_1+rD}{N} + \mu\right)&=&\gamma_0 T_0 + \gamma_1 T_1 + \gamma_d D\\
     R&=&\frac{\gamma_0 T_0 + \gamma_1 T_1 + \gamma_d D}{\rho  \frac{pT_0+qT_1+rD}{N} + \mu}
\end{eqnarray*}
Then,
\begin{eqnarray}\label{R*}
     R^*&=&\frac{\gamma_0 T^*_0 + \gamma_1 T^*_1 + \gamma_d D^*}{\rho  \frac{pT^*_0+qT^*_1+rD^*}{N} + \mu}
\end{eqnarray}

\begin{eqnarray*}
    \frac{dD}{dt}   &=& \delta T_1 - (\mu+\gamma_d)D\\
    D&=&\frac{\delta T_1}{\mu+\gamma_d}
\end{eqnarray*}
Then,
\begin{eqnarray}\label{D*}
    D^*&=&\frac{\delta T^*_1}{\mu+\gamma_d}
\end{eqnarray}

Using \ref{T0*}, \ref{T1*} and \ref{D*}

\begin{align*}
    pT^*_0+qT^*_1+rD^*&=p\cdot \frac{\xi E^*}{\mu+\gamma_0+\omega}+q\cdot \frac{\omega}{\mu+\gamma_1+\delta}\cdot \frac{\xi E^*}{\mu+\gamma_0+\omega}+r\cdot \frac{\delta T^*_1}{\mu+\gamma_d}\\
    &=p\cdot \frac{\xi E^*}{\mu+\gamma_0+\omega}+q\cdot \frac{\omega}{\mu+\gamma_1+\delta}\cdot \frac{\xi E^*}{\mu+\gamma_0+\omega}+r\cdot \frac{\delta}{\mu+\gamma_d}\cdot \frac{\omega}{\mu+\gamma_1+\delta}\cdot \frac{\xi E^*}{\mu+\gamma_0+\omega}\\
\end{align*}
Therefore
\begin{align}\label{uno}
    pT^*_0+qT^*_1+rD^*&=
    &\frac{\xi E^*}{\mu+\gamma_0+\omega}\left(p+\frac{q\omega}{\mu+\gamma_1+\delta}+\frac{r\omega\delta}{(\mu+\gamma_d)(\mu+\gamma_1+\delta)}\right)
\end{align}
By \ref{S*}
\begin{align*}
    S^*=&\displaystyle\frac{\mu N}{\frac{\beta}{N}(pT_0^*+qT_1^*+rD^*)+\mu+\alpha} \\
    =&\displaystyle\frac{\mu N}{\frac{\beta(pT_0^*+qT_1^*+rD^*)+\mu N+\alpha N}{N}}
\end{align*}
then,
\begin{align}\label{S* despej}
    S^*
    =&\frac{\mu N\cdot N}{\beta(pT_0^*+qT_1^*+rD^*)+\mu N+\alpha N}
\end{align}
multiplying by $\frac{1}{N}$
\begin{align*}
 \frac{S^*}{N}=&\frac{1}{N}\cdot \frac{\mu N\cdot N}{\beta(pT_0^*+qT_1^*+rD^*)+\mu N+\alpha N}   \\
 \frac{S^*}{N}=&\frac{\mu N}{\beta(pT_0^*+qT_1^*+rD^*)+\mu N+\alpha N}  
\end{align*}
multiplying by $\beta$
\begin{align*}
    \frac{\beta S^*}{N}=&\frac{\mu \beta N}{\beta(pT_0^*+qT_1^*+rD^*)+\mu N+\alpha N}  
\end{align*}
replacing \ref{uno}
\begin{align}\label{dos}
    \frac{\beta S^*}{N}=&\frac{\mu \beta N}{\beta\cdot\frac{\xi E^*}{\mu+\gamma_0+\omega}\left(p+\frac{q\omega}{\mu+\gamma_1+\delta}+\frac{r\omega\delta}{(\mu+\gamma_d)(\mu+\gamma_1+\delta)}\right)+\mu N+\alpha N}  
\end{align}
now at $E^*$
\begin{align*}
    E^*&=\beta S^*\cdot \frac{pT^*_0+qT^*_1+rD^*}{N}\\
    &=\frac{\beta S^*}{N}(pT^*_0+qT^*_1+rD^*) \hspace{2cm} \text{replacing \ref{dos} and \ref{uno}}\\
    &=\frac{\mu \beta N}{\frac{\beta \xi E^*}{\mu+\gamma_0+\omega}\left(p+\frac{q\omega}{\mu+\gamma_1+\delta}+\frac{r\omega\delta}{(\mu+\gamma_d)(\mu+\gamma_1+\delta)}\right)+\mu N+\alpha N}\cdot E^*\frac{\xi }{\mu+\gamma_0+\omega}\left(p+\frac{q\omega}{\mu+\gamma_1+\delta}+\frac{r\omega\delta}{(\mu+\gamma_d)(\mu+\gamma_1+\delta)}\right)
\end{align*}
if \begin{equation}\label{A}
    A=\frac{\xi}{\mu+\gamma_0+\omega}\left(p+\frac{q\omega}{\mu+\gamma_1+\delta}+\frac{r\omega\delta}{(\mu+\gamma_d)(\mu+\gamma_1+\delta)}\right)
\end{equation}
\begin{align*}
    E^*=&\frac{\mu \beta N}{E^*\beta A+\mu N+\alpha N}\cdot E^* A\\
    E^* (E^*\beta A+\mu N+\alpha N)=&\mu \beta N E^* A\\
    \frac{E^* (E^*\beta A+\mu N+\alpha N)}{E^*}=&\frac{\mu \beta N E^* A}{E^*}\\
    E^*\beta A+\mu N+\alpha N=&\mu \beta N A\\
     E^*\beta A=&\mu \beta N A-\mu N-\alpha N\\
     E^*=&\frac{\mu \beta N A-\mu N-\alpha N}{A\beta}\\
     =&N\left(\frac{\mu \beta A-\mu-\alpha}{A\beta}\right)\\
     =&N\left(\frac{\mu \beta A}{A \beta}-\left(\frac{\mu + \alpha}{A\beta}\right)\right)\\
\end{align*}
therefore,
\begin{align}\label{E desp}
    E^*
     =&N\left(\mu-\left(\frac{\mu + \alpha}{\frac{\beta \xi }{\mu+\gamma_0+\omega}\left(p+\frac{q\omega}{\mu+\gamma_1+\delta}+\frac{r\omega\delta}{(\mu+\gamma_d)(\mu+\gamma_1+\delta)}\right)}\right)\right)
\end{align}
Now, using \ref{S* despej}
\begin{align*}
    S^*
    =&\frac{\mu N\cdot N}{\beta(pT_0^*+qT_1^*+rD^*)+\mu N+\alpha N}
\end{align*}
by \ref{uno} replacing $pT^*_0+qT^*_1+rD^*$ at $S^*$
\begin{align*}
    S^*
    =&\frac{\mu N\cdot N}{\beta\frac{\xi E^*}{\mu+\gamma_0+\omega}\left(p+\frac{q\omega}{\mu+\gamma_1+\delta}+\frac{r\omega\delta}{(\mu+\gamma_d)(\mu+\gamma_1+\delta)}\right)+\mu N+\alpha N}
\end{align*}
using \ref{E desp} and the substitution \ref{A}
\begin{align*}
    S^*=&\frac{\mu N\cdot N}{E^*\beta A+\mu N+\mu N}\\
    =&N\left(\mu-\left(\frac{\mu+\alpha}{A\beta}\right)\right)A\beta +\mu N+\alpha N\\
    =&\frac{\mu N\cdot N}{N\left(\mu-\left(\frac{\mu+\alpha}{A\beta}\right)\right)A \beta +\mu N+\alpha N}\\
    =&\frac{\mu N}{\mu A \beta-\frac{(\mu+\alpha)A\beta}{A\beta}+\mu+\alpha}\\
    =&\frac{\mu N}{\mu A\beta-\mu-\alpha+\mu+\alpha}\\
    =&\frac{\mu N}{\mu A\beta}\\
      =&\frac{ N}{ A\beta}\\
\end{align*}
then,
\begin{align} \label{S* listo}
    S^*=&\frac{N}{\frac{\beta\xi }{\mu+\gamma_0+\omega}\left(p+\frac{q\omega}{\mu+\gamma_1+\delta}+\frac{r\omega\delta}{(\mu+\gamma_d)(\mu+\gamma_1+\delta)}\right)}
\end{align}
using the substitution
\begin{equation}\label{B}
    B=\frac{\beta\xi }{\mu+\gamma_0+\omega}\left(p+\frac{q\omega}{\mu+\gamma_1+\delta}+\frac{r\omega\delta}{(\mu+\gamma_d)(\mu+\gamma_1+\delta)}\right)
\end{equation}
Now using the equation \ref{C*}
\begin{align*}
    C^*&=\displaystyle \frac{\alpha}{\mu}S^*+\frac{\phi}{\mu}E^*
\end{align*}
replacing \ref{S* listo} with the substitution \ref{B} and the equation \ref{E desp}
\begin{align*}
    C^*=&\frac{\alpha}{\mu}\left(\frac{N}{B}\right)+\frac{\phi}{\mu}\left(N\left(\mu-\frac{(\mu+\alpha)}{B}\right)\right)\\
    =&\frac{\alpha}{\mu}\cdot\frac{N}{B}+\frac{\phi}{\mu}\cdot N\left(\frac{B\mu-\mu-\alpha}{B}\right)\\
    =&\frac{N}{\mu B}\left(\alpha+\phi(B\mu-\mu-\alpha)\right)\\
    =&\frac{N}{\mu B}(\alpha+\phi B-\phi \mu-\phi \alpha)\\
    =&\frac{N}{\mu B}(\alpha-\phi \alpha+\phi B-\phi \mu)\\
    =&\frac{N}{\mu B}(\alpha(1-\phi)+\phi (B- \mu))\\
\end{align*}
 
then, retaking \ref{B}
 
 \begin{align*}
     C^*=&\frac{N\left(\alpha(1-\phi)+\phi \left(\frac{\beta\xi }{\mu+\gamma_0+\omega}\left(p+\frac{q\omega}{\mu+\gamma_1+\delta}+\frac{r\omega\delta}{(\mu+\gamma_d)(\mu+\gamma_1+\delta)}\right)- \mu\right)\right)}{\mu \cdot \frac{\beta\xi }{\mu+\gamma_0+\omega}\left(p+\frac{q\omega}{\mu+\gamma_1+\delta}+\frac{r\omega\delta}{(\mu+\gamma_d)(\mu+\gamma_1+\delta)}\right)}
 \end{align*}
using \ref{E desp} with the substitution \ref{B} at \ref{T0*}, 
\begin{align*}
   T^*_0&=\frac{\xi E^*}{\mu+\gamma_0+\omega} \\
   &=\frac{\xi}{\mu+\gamma_0+\omega}\left(N\left(\mu-\left(\frac{\mu + \alpha}{\frac{\beta \xi }{\mu+\gamma_0+\omega}\left(p+\frac{q\omega}{\mu+\gamma_1+\delta}+\frac{r\omega\delta}{(\mu+\gamma_d)(\mu+\gamma_1+\delta)}\right)}\right)\right)\right)\\
   &=\frac{N \xi}{\mu+\gamma_0+\omega}\left(\frac{B\mu-\mu-\alpha}{B}\right)\\
   &=N \cdot \frac{\xi}{\mu+\gamma_{0}+\omega}\left(\frac{B\mu-\mu-\alpha}{\frac{\beta \xi}{\mu+\gamma_{0}+\omega}\left(p+\frac{q\omega}{\mu+\gamma_{1}+\delta}+\frac{r\omega \delta}{(\mu+\gamma_d)(\mu+\gamma_{1}+\delta)}\right)}\right)\\
   &=N \left(\frac{B\mu-\mu-\alpha}{\beta \left(p+\frac{q\omega}{\mu+\gamma_{1}+\delta}+\frac{r\omega \delta}{(\mu+\gamma_d)(\mu+\gamma_{1}+\delta)}\right)}\right)\\
   &=\frac{NB\mu}{\beta \left(p+\frac{q\omega}{\mu+\gamma_{1}+\delta}+\frac{r\omega \delta}{(\mu+\gamma_d)(\mu+\gamma_{1}+\delta)}\right)}-\frac{N(\mu+\alpha)}{\beta \left(p+\frac{q\omega}{\mu+\gamma_{1}+\delta}+\frac{r\omega \delta}{(\mu+\gamma_d)(\mu+\gamma_{1}+\delta)}\right)}\\
   & = \frac{N\beta\frac{\xi}{\mu+\gamma_0+\omega}\left(p+\frac{q\omega}{\mu+\gamma_{1}+\delta}+\frac{r\omega \delta}{(\mu+\gamma_d)(\mu+\gamma_{1}+\delta)}\right)\mu}{\beta \left(p+\frac{q\omega}{\mu+\gamma_{1}+\delta}+\frac{r\omega \delta}{(\mu+\gamma_d)(\mu+\gamma_{1}+\delta)}\right)}-\frac{N(\mu+\alpha)}{\beta \left(p+\frac{q\omega}{\mu+\gamma_{1}+\delta}+\frac{r\omega \delta}{(\mu+\gamma_d)(\mu+\gamma_{1}+\delta)}\right)}\\
   & = \frac{N\xi\mu}{\mu+\gamma_0+\omega}-\frac{N(\mu+\alpha)}{\beta \left(p+\frac{q\omega}{\mu+\gamma_{1}+\delta}+\frac{r\omega \delta}{(\mu+\gamma_d)(\mu+\gamma_{1}+\delta)}\right)}
\end{align*}
then,
\begin{align}\label{T*0 desp}
   T^*_0&=\frac{N\xi\mu}{\mu+\gamma_0+\omega}-\frac{N(\mu+\alpha)}{\beta \left(p+\frac{q\omega}{\mu+\gamma_{1}+\delta}+\frac{r\omega \delta}{(\mu+\gamma_d)(\mu+\gamma_{1}+\delta)}\right)}
\end{align}
when $\rho=0$ at \ref{T1*} and using \ref{T*0 desp}
\begin{align*}
  T^*_1&=\frac{\omega}{\mu+\gamma_1+\delta}T^*_0 
\end{align*}
\begin{align}
   T^*_1&=\frac{N\omega}{\mu+\gamma_1+\delta}\left(\frac{\xi\mu}{\mu+\gamma_0+\omega}-\frac{(\mu+\alpha)}{\beta \left(p+\frac{q\omega}{\mu+\gamma_{1}+\delta}+\frac{r\omega \delta}{(\mu+\gamma_d)(\mu+\gamma_{1}+\delta)}\right)}\right)
\end{align}
for $D^*$ by \ref{D*}
\begin{align*}
   D^*&=\frac{\delta T^*_1}{\mu+\gamma_d} \\
   &=\frac{\gamma_1}{\mu+\gamma_d}\cdot \frac{N\omega}{\mu+\gamma_1+\delta}\left(\frac{\xi\mu}{\mu+\gamma_0+\omega}-\frac{(\mu+\alpha)}{\beta \left(p+\frac{q\omega}{\mu+\gamma_{1}+\delta}+\frac{r\omega \delta}{(\mu+\gamma_d)(\mu+\gamma_{1}+\delta)}\right)}\right) \\
   & = \frac{ \omega N}{(\mu+\gamma_d)(\mu+\gamma_1+\delta)}\left(\frac{\xi\mu}{\mu+\gamma_0+\omega}-\frac{(\mu+\alpha)}{\beta \left(p+\frac{q\omega}{\mu+\gamma_{1}+\delta}+\frac{r\omega \delta}{(\mu+\gamma_d)(\mu+\gamma_{1}+\delta)}\right)}\right)
\end{align*}

\section*{Acknowledgments}
The authors are grateful for the support of the Research Center in Pure and Applied Mathematics and the Department of Mathematics at the University of Costa Rica.

\bibliographystyle{siamplain}

\end{document}

%% file: ex_shared.tex
\usepackage{lipsum}
\usepackage{amsfonts}
\usepackage{graphicx}
\usepackage{epstopdf}
\usepackage{algorithmic}
\usepackage{xurl}
\ifpdf
  \DeclareGraphicsExtensions{.eps,.pdf,.png,.jpg}
\else
  \DeclareGraphicsExtensions{.eps}
\fi

\newsiamremark{remark}{Remark}
\newsiamremark{hypothesis}{Hypothesis}
\crefname{hypothesis}{Hypothesis}{Hypotheses}
\newsiamthm{claim}{Claim}

\headers{The influence of trap music on adolescents}{Rodríguez, G. Garita, T. and Sanchez, F.}

\title{A Dynamical Systems Analysis of Trap Music Exposure and School Dropout Among Costa Rican Adolescents}

\author{Geraldine Rodríguez\thanks{\email{geraldine.rodriguez\_r@ucr.ac.cr}}
\and Tonny A. Garita\thanks{\email{tonny@newsummitacademy.com}}\and Fabio S\'anchez\thanks{\email{fabio.sanchez@ucr.ac.cr}}}

\usepackage{amsopn}

\makeatletter
\newcommand*{\addFileDependency}[1]{
  \typeout{(#1)}
  \@addtofilelist{#1}
  \IfFileExists{#1}{}{\typeout{No file #1.}}
}
\makeatother

\newcommand*{\myexternaldocument}[1]{%
    \externaldocument{#1}%
    \addFileDependency{#1.tex}%
    \addFileDependency{#1.aux}%
}